\documentclass[aps,pra,showpacs,raggedbottom, 
amssymb,twocolumn,groupedaddress]{revtex4-1}
\usepackage{graphicx}
\usepackage{amsmath}
\usepackage{amsfonts}
\usepackage{amssymb}
\usepackage{epsfig,libertine}
\usepackage[libertine]{newtxmath}
\usepackage{color}
\usepackage{dsfont, hyperref, xcolor}
\usepackage{comment}
\usepackage{enumerate}

\newcommand{\abs}[1]{\left\vert#1\right\vert}

\newcommand{\Tr}[1]{\text{Tr}\left\{#1\right\}}
\newcommand{\ParTr}[2]{\text{Tr}_{#1}\left\{#2\right\}}
\newcommand{\bra}[1]{\left\langle#1\right\vert}
\newcommand{\ket}[1]{\left\vert#1\right\rangle}

\begin{document}

\title{Causal games of work extraction with indefinite causal order}

\author{Gianluca~Francica}
\email{gianluca.francica@gmail.com}

\address{Dipartimento di Fisica e Astronomia ``G. Galilei'', Universit\`{a} degli Studi di Padova, via Marzolo 8, 35131 Padova, Italy}

\date{\today}

\begin{abstract}
An indefinite causal order, where the causes of events are not necessarily in past events, is predicted by the process matrix framework. A fundamental question is how these non-separable causal structures can be related to the thermodynamic phenomena. Here, we approach this problem by considering the existence of two cooperating local Maxwell's  demons which try to exploit the presence of global correlations and indefinite causal order to optimize the extraction of work.
Thus, we prove that it is possible to have a larger probability to lower the local energy to zero if causal inequalities are violated, and that can be extracted more average work with respect to a definite causal order. However, for non-interacting parties, for the system considered the work extractable cannot be larger than the definite causal order bound.
\end{abstract}

\maketitle

\section{Introduction}
In the common view of the nature, the events have a definite causal order, i.e., the causes of events are in past events. In contrast, it is known that the process matrix formalism of Oreshkov, Costa and Brukner~\cite{Oreshkov12} allows causal structures compatible with local quantum mechanics for two parties Alice and Bob, which are causally non-separable, i.e., neither Alice comes before Bob nor Bob comes before Alice, nor a mixture thereof. This indefinite causal order can be exploited in some ``causal games'', e.g., allows us to find a strategy violating the so-called causal inequalities~\cite{Oreshkov12,Branciard16}, even if there also are causally non-separable processes which admit a causal model~\cite{Feix16}, i.e., they do not violate causal inequalities (a popular example is the quantum switch model~\cite{Chiribella13}).
Concerning the question if there is causation in fundamental physics, some relevant insights are recently given in Ref.~\cite{Adlam22}. A further fundamental question, which here we take in exam, can be if indefinite causal order can be compatible with the second law of thermodynamics.
Furthermore, concerning the applications, we note that indefinite causal order can provide advantages in certain communication~\cite{feix15,guerin16,ebler18} and computing tasks~\cite{chiribella12,araujo14}, and recent experiments have been performed~\cite{procopio15,rubino17,goswami18,goswami20,rubino21}.
Recently, a certain attention has also been paid to understand the role of indefinite causal order to achieve thermodynamic tasks~\cite{Nie20,Cao21,Felce21,Felce20,Guha20,Simonov22,Dieguez22,Guha22,Capela22}.
We note that the  daemonic ergotropy~\cite{Francica17}, where the information gained by performing measurements on a part of the system can be communicated to locally extract work from a different part, can be naturally related to a causal structure.
Basically, in the daemonic ergotropy extraction Alice and Bob share a correlated state, if Alice performs measurements on her part and communicates to Bob the outcomes, Bob can perform an optimal unitary cycle to extract the maximum work from his part. Of course, the measurements performed by Alice are in the past of Bob, thus the protocol exhibits a definite causal order. Here, we adopt the same point of view, with the addition that the roles of Alice and Bob can be also reversed, in order to get a causally non-separable structure. As we will show, this allows us to get a work extraction where indefinite causal order plays some major role.

\section{Preliminaries of thermodynamics}
Let us introduce some concepts of work extraction in finite quantum systems. We are interested in extracting work from a quantum system in an initial state $\rho_0$ by using a cyclic unitary transformation, i.e., by considering a unitary time evolution operator $U$ generated by a time-dependent Hamiltonian which at the final time is equal to the initial Hamiltonian $H$. We aim to maximize the average work extracted from the system, or equivalently to minimize the final average energy $\langle E \rangle = \Tr{H \rho}$, where $\rho=U\rho_0 U^\dagger$ is the final state, over the set of all the unitary operators $U$, since in general the average work extracted is equal to minus the change of average energy if the system is thermally isolated. As shown in Ref.~\cite{allahverdyan04}, the final average energy $\langle E \rangle$ is minimal if the final state is a passive state, i.e., $\rho=  P_{\rho_0} \equiv \sum_k r_k \ket{\epsilon_k} \bra{\epsilon_k}$, where we have ordered the labels of eigenstates of $H$ and of $\rho_0$ such that $H=\sum_k \epsilon_k \ket{\epsilon_k} \bra{\epsilon_k}$, with $\epsilon_k\leq \epsilon_{k+1}$, and $\rho_0=\sum_k r_k \ket{r_k}\bra{r_k}$, with $r_k\geq r_{k+1}$. The resulting average work is known as ergotropy. In particular, for an initial pure state, the passive state will be the ground-state $\ket{\epsilon_1}$.
Concerning the local extraction of work, if the system consists of two parties, Alice and Bob, which do not interact with each other, Alice can perform measurements on her part, described by a set of orthogonal projectors $\Pi^A_x$ of rank one, and can communicate the outcomes $x$ to Bob. Thus, the state of Bob collapses into $\rho_{B|x}=\ParTr{A}{\Pi^A_x \otimes I^B \rho_0}/p(x)$ with probability $p(x)=\Tr{\Pi^A_x \otimes I^B \rho_0}$, where $I^B$ is the identity matrix on the Bob's Hilbert space, and Bob can perform a local cyclic unitary transformation $U_x$ conditioned on the outcome $x$ of the measurement. We aim to maximize the average work extracted from the Bob's part, or equivalently to minimize the final average energy of Bob $\langle E_B \rangle = \sum_x p(x)\Tr{H_B U_x\rho_{B|x} U_x^\dagger}$ over the set of all the unitary operators $U_x$ for given projectors $\Pi^A_x$, where $H_B$ is the Hamiltonian of Bob. The resulting average work extracted by Bob, known as daemonic ergotropy, is minus the change of the average energy of Bob and has been investigated in terms of the correlations between the two parties in Ref.~\cite{Francica17}. By considering that the final average energy of Alice is $\langle E_A \rangle = \sum_x \Tr{ \rho_0 \Pi^A_x  H_A \Pi^A_x\otimes I^B}$, where $H_A$ is the Hamiltonian of Alice, the final average energy of the total system is $\langle E \rangle = \langle E_A \rangle + \langle E_B \rangle$.
Typically, the average energy of Alice can change because of the measurements, however for the special case where the Hilbert spaces of Alice and Bob have the same dimension $d$ and $\rho_0$ is a maximally entangled pure state, the reduced states of Alice and Bob are equal to the completely mixed state, so that the average energy of Alice does not change for any measurements and the daemonic ergotropy, i.e., the average work extracted $\langle w \rangle$ is  minus the change of average energy of the total system. In particular, for any projective measurements $\Pi^A_x$, we get $\langle w \rangle = \sum_i \epsilon^B_i /d - \epsilon^B_1$, where $\epsilon^B_i$ are the eigenvalues of $H_B$ and $\epsilon^B_1$ is the lowest one, because Bob can perform local cyclic unitary transformations $U_x$ to always achieve the state with lowest energy.
Of course, in this scheme of work extraction, we have a definite causal order. In the next section, with the aim to consider causally non-separable structures, we will modify this scheme.
We will be interested in the average work extracted and so in the final average energy $\langle E \rangle$ of the total system, which, given the final state $\rho$, can be calculated as $\langle E \rangle = \Tr{\rho H}$. Moreover, we will be interested in a certain subsystem $X$, having final average energy $\langle E_X \rangle = \Tr{\rho_X H_X}$, where $\rho_X$ and $H_X$ are the final reduced state and the Hamiltonian of the subsystem, respectively. In particular, we note that in general the final energy of the subsystem $X$ has the probability distribution $p(E_X) = \sum_i \bra{\epsilon^X_i}\rho_X \ket{\epsilon^X_i} \delta(E_X - \epsilon^X_i)$, where $\ket{\epsilon^X_i}$ is the eigenstate of $H_X$ with eigenvalue $\epsilon^X_i$, so that $\langle E_X \rangle = \int E_X p(E_X) dE_X$.

\section{Causal games of work extraction}
\begin{figure}
[h!]
\centering
\includegraphics[width=0.65\columnwidth]{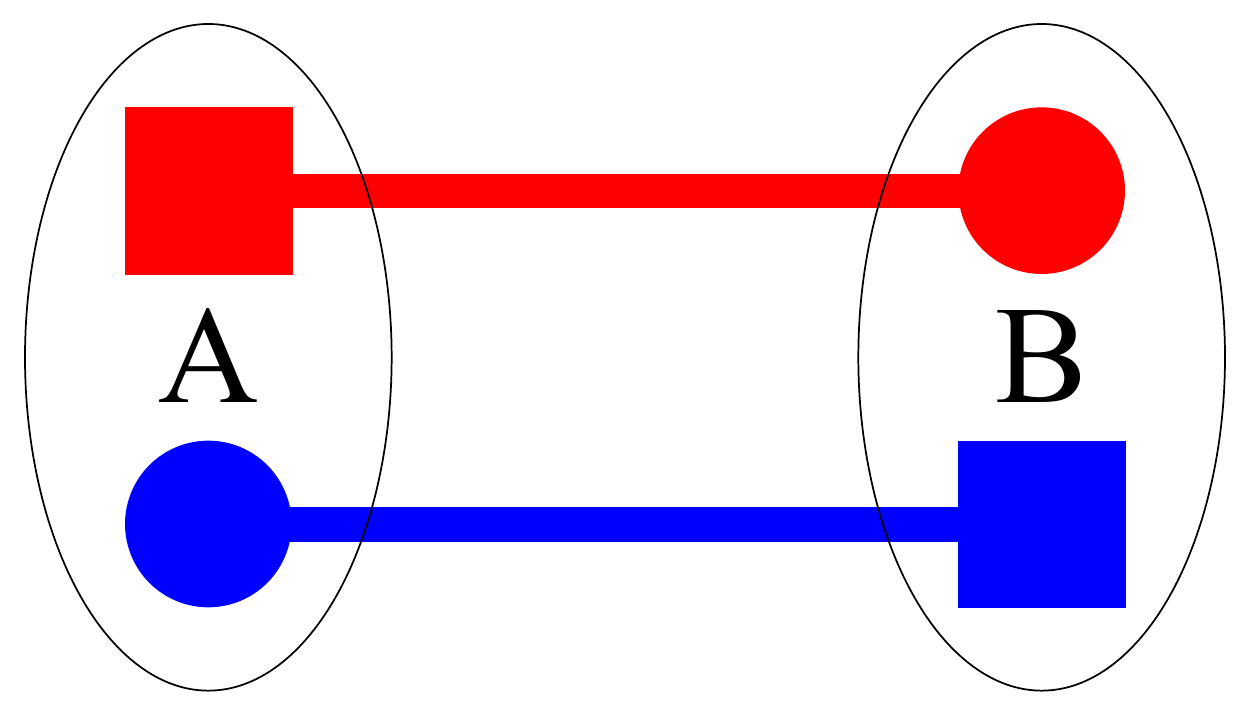}
\caption{The system is made of four parties, represented by two squares and two circles. The parties in $A$ and $B$ belong to Alice and Bob, respectively, and the parties connected by a line are initially correlated. Local measurements and unitary operations are performed on the squares and the circles, respectively.
}
\label{fig:scheme}
\end{figure}

To link the daemonic ergotropy scheme to an indefinite causal order structure, we assume that Alice and Bob share two correlated states. In particular both Alice and Bob have a square qubit and a circle qubit. The square qubit of Alice and the circle one of Bob (red qubits) are correlated, in particular are in a singlet state, and the circle qubit of Alice and the square one of Bob (blue qubits) are also in a singlet state. Alice and Bob can perform local operations and can communicate each other. In detail, Alice and Bob perform measurements on the square qubits, and unitary operations on the circle ones. A schematic illustration of the system is given in Fig.~\ref{fig:scheme}.
Since Alice and Bob can communicate, due to the presence of global correlations, in principle they can perform optimized unitary operations depending on the measurements to achieve a precise task. Here, we consider the possibility to have causally non-separable processes, and a thermodynamic task that is the work extraction from the qubits system, i.e., from the two singlet states. What is the role of indefinite causal order in achieving this task? Due to the violation of causal inequalities, in particular, the probability to guess the neighbor values of the measurement is larger in the causally non-separable case~\cite{Branciard16}, one can expect to extract more work than in the causally separable case. However, the answer is not that simple, let us show why. We start to define the Hamiltonian of the system. The qubits do not interact each other, and a qubit has the Hamiltonian $H_{q}=\epsilon (\sigma_x+1)/2$, with $\epsilon>0$, where $\sigma_\alpha$, with $\alpha=x,y,z$, are the Pauli matrices. Alice and Bob will perform measurements of $\sigma_z$ on their square qubits (the main results do not change by changing basis). If the system is in the initial state, Alice gets the bit $x$ and the singlet state collapses in $\ket{x}\otimes\ket{\bar{x}}$ with probability $p(x)=1/2$, where $x=0,1$, $\ket{x}$ is eigenstate of $\sigma_z$ with eigenvalue $2x-1$ and $\bar{x}=0$ if $x=1$, $\bar{x}=1$ if $x=0$. Similarly, Bob gets the bit $y$ and the singlet state collapses in $\ket{y}\otimes\ket{\bar y}$ with probability $p(y)=1/2$. We note that one measurement does not change the  average energy of a singlet state which is $\epsilon$ for any value of the bit obtained.
This energy can be lowered due to a local cyclic unitary transformation.
We start to take in exam a definite causal order where Alice precedes Bob, i.e., $A\prec B$, so that, at first, Alice performs a measurement on her square red qubit, and communicate to Bob the bit $x$, and performs a local cyclic unitary transformation $U_A$ on her circle blue qubit. Later, Bob uses the bit $x$ of information received to perform a local cyclic unitary transformation $U_x$ on his circle red qubit, so that the initial red singlet state becomes $\ket{x}\otimes U_x \ket{\bar x}$ with probability  $p(x)=1/2$, thus with the aim to lower the energy, Bob performs the unitary $U_x$ such that $U_x \ket{\bar x}=\ket{-}$, where $\ket{\pm}$ is the eigenstate of $\sigma_x$ with eigenvalue $\pm 1$, and the final average energy of the two qubits will be $\epsilon/2$ with equal probability, so that its average is $\langle E_{red}\rangle = \epsilon/2$. In detail, the local unitary $U_x$ is such that $U_0\ket{1}=\ket{-}$, $U_0\ket{0}=\ket{+}$, $U_1\ket{0}=\ket{-}$ and $U_1\ket{1}=\ket{+}$. The blue qubits are in the state $(U_A\ket{0}\otimes \ket{1} - U_A\ket{1}\otimes \ket{0})/\sqrt{2}$ with an average energy equal to $\epsilon$ for any $U_A$. If Bob performs a measurement on his square blue qubit, we get the state $U_A\ket{\bar{y}}\otimes\ket{y}$ with probability $p(y)=1/2$, but in average the energy remains $\langle E_{blue}\rangle = \epsilon$. Thus, in this case the final average energy of the four qubits is equal to $\langle E\rangle=\langle E_{red}\rangle+\langle E_{blue}\rangle=\epsilon/2 + \epsilon= 3\epsilon/2$, which is the lowest value gettable for a definite causal order.
Since the measurements do not change the average energy, the average work extracted is $\langle w \rangle = 2\epsilon - \langle E \rangle$ and gets the maximum value $\langle w \rangle = \epsilon/2$, which is also equal to the daemonic ergotropy of the red qubits.
Of course, if Alice and Bob do not communicate but only use local operations, in average they cannot lower the energy of the singlet states, so that the average work is zero, $\langle w\rangle =0$, in agreement with the second law of thermodynamics. Thus, the gain achieved is related to the use of the information acquired.

In contrast, in the presence of indefinite causal order Alice and Bob are given the bit inputs $x$ and $y$, with probability $p(x,y)$, and return the bit outputs $a$ and $b$, respectively.
By taking in account Alice, for each input $x$ and output $a$, we associate an operation described by a completely positive map $\mathcal M^{A_I A_O}_{a|x}: \mathcal L (\mathcal H^{A_I})\to \mathcal L (\mathcal H^{A_O})$, where $\mathcal L (\mathcal H^{X})$ is the space of linear operators over the Hilbert space $\mathcal H^X$ of dimension $d_X=2$. We note that all the maps must sum up to a trace-preserving map.
Using the Choi-Jamiol{\l}kowski isomorphism~\cite{Choi75,Jamiolkowski72}, we represent the map $\mathcal M^{A_I A_O}_{a|x}$ as the operator $M^{A_I A_O}_{a|x}=[I^{A_I}\otimes \mathcal M^{A_I A_O}_{a|x}(\ket{\varphi^+}\bra{\varphi^+})]^T \in \mathcal L (\mathcal H^{A_I} \otimes \mathcal H^{A_O})$, where $I^{X}$ is the identity matrix on $\mathcal H^{X}$ and $\ket{\varphi^+}=\sum_i \ket{ii}$. The operators $M^{A_I A_O}_{a|x}$ are such that $M^{A_I A_O}_{a|x}\geq 0$ for each $a$ and $\ParTr{A_O}{\sum_a M^{A_I A_O}_{a|x}}=I^{A_I}$.
Similarly, for Bob we get the operators $M^{B_I B_O}_{b|y}\in \mathcal L (\mathcal H^{B_I} \otimes \mathcal H^{B_O})$. The joint conditional probability reads
\begin{equation}\label{cond_prob}
p(a,b|x,y)=\Tr{(M^{A_I A_O}_{a|x}\otimes M^{B_I B_O}_{b|y})W}\,,
\end{equation}
where $W\in \mathcal L (\mathcal H^{A_I} \otimes \mathcal H^{A_O}\otimes\mathcal H^{B_I} \otimes \mathcal H^{B_O})$ is the so-called process matrix, which is an hermitian operator such that the probabilities given by Eq.~\eqref{cond_prob} are non-negative and normalized.
In particular, a process matrix $W$ needs to satisfy the conditions~\cite{Araujo15}
\begin{eqnarray}
\label{eq1} W&\geq&0\,,\\
\label{eq2} \Tr{W} &=& d_{A_O} d_{B_O}\,,\\
\label{eq3} _{B_I B_O} W &=& _{A_O B_I B_O} W\,,\\
\label{eq4} _{A_I A_O} W &=& _{A_I A_O B_O} W\,,\\
\label{eq5} W &=& _{B_O} W+{_{A_O} W} -{_{A_O B_O}W}\,,
\end{eqnarray}
where we have defined the operation
\begin{equation}
_{X} W = \frac{I^X}{d_X}\otimes\ParTr{X}{W}\,.
\end{equation}
If Bob cannot signal to Alice or Alice cannot signal to Bob we have the process matrices $W^{A\prec B}=W^{A_I A_O B_I}\otimes I^{B_O}$ or $W^{B\prec A}=W^{A_I B_I B_O}\otimes I^{A_O}$, respectively, where  $W^{A_I A_O B_I}\in \mathcal L (\mathcal H^{A_I}\otimes \mathcal H^{A_O}\otimes \mathcal H^{B_I})$ and $W^{A_I B_I B_O}\in \mathcal L (\mathcal H^{A_I}\otimes \mathcal H^{B_I}\otimes \mathcal H^{B_O})$. In detail, from Eqs.~(\ref{eq1}-\ref{eq5}) we get the conditions for the matrix $W^{A_I A_O B_I}$
\begin{eqnarray}
W^{A_I A_O B_I} &\geq& 0 \,,\\
\Tr{W^{A_I A_O B_I}} &=& d_{A_O}\,,\\
_{B_I} W^{A_I A_O B_I} &=& _{A_O B_I} W^{A_I A_O B_I}\,,
\end{eqnarray}
and similar conditions for the matrix $W^{A_I B_I B_O}$.
Therefore, a process is causally separable if the process matrix can be expressed in a convex combination
\begin{equation}\label{eq W sep}
W_{sep}=q W^{A\prec B} + (1-q) W^{B\prec A}\,.
\end{equation}
We recall that causal non-separability can be inferred by using causal inequalities~\cite{Oreshkov12,Branciard16,Araujo15,Baumeler14,Oreshkov16}. An example of causal inequality is a bound of the probability of success of the ``guess your neighbor's input'' game~\cite{Branciard16}. For uniform input bits $x$ and $y$, the probability of success is $p_{succ}=1/4\sum_{x,y}p(a=y,b=x|x,y)$ and for a separable process $p_{succ}\leq 1/2$, but it is known that there are causally non-separable processes such that $p_{succ}>1/2$. Anyway, there also are non-separable process, having a causal model, that do not violate causal inequalities~\cite{Feix16}.
One can expect that $p_{succ}$ will play a role in the work extraction. The work extraction scheme can be generalized to the case of indefinite causal order by requiring that Alice and Bob perform the local cyclic unitary transformations $U_a$ and $U_b$, respectively, which depend on the bits $a$ and $b$.
It is easy to see that the final state is the mixture
\begin{eqnarray}
\nonumber \rho &=&  \sum_{x,y,a,b} p(x,y) p(a,b|x,y) \ket{x}\bra{x}\otimes U_b \ket{\bar{x}}\bra{\bar{x}}U_b^\dagger\\
\label{eq rho}&&\otimes  U_a \ket{\bar{y}}\bra{\bar{y}}U_a^\dagger \otimes \ket{y}\bra{y}\,,
\end{eqnarray}
where in our case $p(x,y)=1/4$.
Of course the two initial non-correlated singlets become correlated but are still separable. The final average energy of the state $\rho$ is
\begin{equation}\label{eq ave}
\langle E \rangle =  \epsilon + p_1 \epsilon + 2 p_2 \epsilon = 2 \epsilon - (p_{succ} - p_2)\epsilon\,,
\end{equation}
where we have defined the probabilities $p_1$ and $p_2$ to wrong one and two bits, respectively, which are $p_2=1/4\sum_{x,y}p(a=\bar{y},b=\bar{x}|x,y)$ and $p_1=1-p_{succ}-p_2$.
To derive Eq.~\eqref{eq ave}, we observe that the final state $\rho$ can be expressed as
\begin{eqnarray}
\nonumber \rho &=& \frac{1}{4} \sum_{x,y}\big ( p(a=y,b=x|x,y) P(\ket{x}\otimes\ket{-}\otimes\ket{-} \otimes\ket{y})\\
\nonumber &&+p(a=y,b=\bar{x}|x,y) P(\ket{x}\otimes\ket{+}\otimes\ket{-}\otimes\ket{y})\\
\nonumber &&+p(a=\bar{y},b=x|x,y) P(\ket{x}\otimes\ket{-}\otimes\ket{+}\otimes\ket{y})\\
&&+p(a=\bar{y},b=\bar{x}|x,y) P(\ket{x}\otimes\ket{+}\otimes\ket{+}\otimes\ket{y})\big)\,,
\end{eqnarray}
where for brevity, given a state $\ket{\psi}$ we have defined the projector $P(\ket{\psi})=\ket{\psi}\bra{\psi}$. Thus, by noting that $\Tr{H_q \ket{x}\bra{x}}=\epsilon/2$, with $x=0,1$ and $\Tr{H_q \ket{\pm}\bra{\pm}}=(1\pm 1)\epsilon/2$, from $\langle E \rangle = \Tr{\rho H}$ we get Eq.~\eqref{eq ave}.
Similarly, we deduce that the final energy of the circle qubits has the probability distribution
\begin{equation}\label{eq prob}
p(E_c)=\sum_{i=0}^2 p_i \delta(E_c- i \epsilon)\,,
\end{equation}
where $p_0=p_{succ}$.
For a definite causal order, the optimal process described above gives $p_{succ}=1/2$, and no chance of getting both bits wrong, thus $p_2=0$ and $p_1=1/2$. In this case, the energy of the circle qubits is $\epsilon$ or zero with the same probability. For an indefinite causal order such that $p_{succ}>1/2$, the probability that this energy is zero is larger than any definite causal order.
Our next question is if it is possible to achieve a gain in the average extracted work, i.e., to get $\langle E \rangle < 3\epsilon/2$. Surprisingly, for the system under consideration, the answer is negative, because of the presence of a non-zero probability $p_2$. In particular, we find the upper bound (see Appendix~\ref{appendix} for the proof)
\begin{equation}\label{eq bb}
p_{succ}-p_2\leq 1/2\,,
\end{equation}
from which $\langle E \rangle\geq 3\epsilon/2$.
Thus, in the presence of indefinite causal order, if $p_{succ}>1/2$ then $p_2 > 0$ and $p_1<1/2$. Concerning the energy of the circle qubits, with respect to the optimal causally separable process, we get a larger probability $p_{succ}$ that is zero, a smaller probability $p_1$ that is $\epsilon$, but the non-zero probability $p_2$ that is $2\epsilon$ gives an average extracted work $\langle w \rangle = 2\epsilon-\langle E \rangle$ not larger than $\epsilon/2$, and so also a larger variance. We note that this result is quite general, i.e., it is not possible to get a gain in the average work extracted. In general, by performing a measurement of the spin with respect to an arbitrary direction, the average energy of the square qubits does not change, and the energy of the circle qubits still has the probability distribution of Eq.~\eqref{eq prob} for optimal local cyclic unitary transformations.
However, if interactions between the parties are allowed, in particular if we consider the interaction between the circle qubits $H_{int}=-\epsilon \ket{+}\bra{+}\otimes\ket{+}\bra{+}$, then the final energy of the circle qubits has the probability distribution
\begin{equation}\label{eq prob}
p(E_c)=p_{succ} \delta(E_c)+(1-p_{succ})\delta(E_c-\epsilon)\,,
\end{equation}
which depends only on the probability $p_{succ}$. In this case, since the average energy of the square qubits does not change, by considering that the interaction lowers the initial average energy to $2\epsilon - \epsilon/4$, the average work extracted is $\langle w \rangle = (p_{succ}-1/4)\epsilon$, which of course, in the presence of an indefinite causal order such that $p_{succ}>1/2$, is larger than the causal bound $\epsilon/4$.
We note that if the qubits are realized by using spinless fermions, the Hamiltonian reads $H=H_{red}+H_{blue}+H_{int}$, where $H_{\alpha}=\epsilon n_\alpha^A + \epsilon n_\alpha^B$ and $H_{int}=-\epsilon n_{blue}^A n_{red}^B$, where $n_\alpha^A$ and $n_\alpha^B$ are the number operators of the fermions of Alice and Bob, respectively, with $\alpha=red,blue$.
Of course, to implement the local cyclic unitary transformations the interaction $H_{int}$ is switched off in the corresponding time interval, and switched on at the end.

In the end, it is worth observing that the work extraction can be related to the formation of correlations between the red qubits and the blue ones in the final state $\rho$ of Eq.~\eqref{eq rho}. In particular, for the optimal process with definite causal order giving $p_{succ}=1/2$, in the final state the red qubits are not correlated with the blue ones. In general, the total correlations are quantified by the mutual information $I_{red:blue}=S_{red}+S_{blue}-S_{red, blue}$, where $S_{red}$ and $S_{blue}$ are the von-Neumann entropies of the reduced final states of the red and blue qubits, respectively. In detail, the von-Neumann entropy of a state $\rho$ is defined as $S=-\Tr{\rho\log_2\rho}$. Conversely, $S_{red, blue}$ is the von-Neumann entropy of the final state $\rho$ of the total system, which explicitly reads $S_{red, blue}=-\sum_{a,b,x,y} \lambda_{abxy}\log_2 \lambda_{abxy}$, where $\lambda_{abxy}= p(x,y) p(a,b|x,y)$ are the eigenvalues of the density matrix $\rho$. Since $\sum_{a,b}\lambda_{abxy} = p(x,y)=1/4$ and $\sum_{x,y} \lambda_{abxy}= p'(a,b)$, where $p'(a,b)$ is the probability to get the outputs $a$ and $b$, we get $S_{red, blue} = 2 + H_{A,B} - I_{I:O}$, where $H_{A,B} = -\sum_{a,b} p'(a,b)\log_2 p'(a,b)$ is the Shannon entropy corresponding to the outputs $a$, $b$, and $I_{I:O}$ is the mutual information between the inputs $x$,$y$ and the outputs $a$, $b$. Then, we get
\begin{equation}\label{eq b0}
I_{red:blue}=S_{red}+S_{blue} -2 + I_{I:O}- H_{A,B}\,,
\end{equation}
from which, since $S_{red}\leq 2$ and $S_{blue}\leq 2$, we get the upper bound
\begin{equation}\label{eq bI}
I_{red:blue} \leq 2 + I_{I:O}- H_{A,B}\,,
\end{equation}
where $0\leq H_{A,B} - I_{I:O}\leq 2$. We note that Eq.~\eqref{eq bI} connects the formation of correlations between the red and blue qubits only to the information exchange.
Concerning $H_{A,B}$, as shown in Ref.~\cite{Francica22}, there exist causally non-separable process matrices $W$ such that $H_{A,B}(W) > \max_{W_{sep}\in \mathcal S_W} H_{A,B}(W_{sep})$, where  $\mathcal S_W$ is the set of all the separable processes $W_{sep}$, defined by Eq.~\eqref{eq W sep}, such that $\Delta(W^{A\prec B})=\Delta(W)$ and $\Delta(W^{B\prec A})=\Delta(W)$, and $\Delta(W)$ is the non-signalling part of the process matrix $W$ defined as $\Delta(W)= {_{A_O B_O}W}$.
However, this does not rule out that $H_{A,B}-I_{I:O}$ is smaller for an indefinite causal order, getting a larger amount of correlations between the red and the blue qubits.
For instance, we consider the process matrix
\begin{equation}
W=\frac{1}{4}\left(I^{\otimes 4} + \alpha(\sigma_z^{A_I}\sigma_z^{A_O}\sigma_z^{B_I}I^{B_O}+\sigma_z^{A_I}I^{A_O}\sigma_x^{B_I}\sigma_x^{B_O})\right)\,,
\end{equation}
with $0\leq \alpha\leq 1/\sqrt{2}$, where the tensor products are implicit, and the local operations
\begin{eqnarray}
M^{A_I A_O}_{0|0}&=&M^{B_I B_O}_{0|0}=0\,,\\
M^{A_I A_O}_{1|0}&=&M^{B_I B_O}_{1|0}=\ket{\varphi^+}\bra{\varphi^+}\,,\\
M^{A_I A_O}_{0|1}&=&M^{B_I B_O}_{0|1}=\ket{0}\bra{0}\otimes \ket{0}\bra{0}\,,\\
M^{A_I A_O}_{1|1}&=&M^{B_I B_O}_{1|1}=\ket{1}\bra{1}\otimes \ket{0}\bra{0}\,.
\end{eqnarray}
We get $p_{succ}=5(1+\alpha)/16$, thus the violation of causal inequalities becomes larger as $\alpha$ increases, for $\alpha=1/\sqrt{2}$ is maximum and we get the causally non-separable process matrix of Ref.~\cite{Branciard16}. We find that $H_{A,B}$ increases,  $H_{A,B}-I_{I:O}$ decreases and $I_{red:blue}$ increases as $\alpha$ increases. In particular, for $\alpha=1/\sqrt{2}$, we get a non-zero mutual information $I_{red:blue}\approx 1.0951$. Moreover, the bound is $I_{I:O}-H_{A,B}+2\approx 1.2993$, so that $I_{red:blue}$ is close to the upper bound since $S_{red}=S_{blue}\approx1.8979$ is close to two. Furthermore, we get $p_2=(5+\alpha)/16$, and so $p_{succ}-p_2=\alpha/4$, in agreement with Eq.~\eqref{eq bb}.
We note that the bound of Eq.~\eqref{eq bb} can be saturated by choosing opportunely the local operations, e.g., for $\alpha=1/\sqrt{2}$ we get $p_{succ}=1/2$ and $p_2=0$ for the operations
\begin{eqnarray}
M^{A_I A_O}_{0|0}&=&\frac{1}{4}\left( I^{\otimes 2} + \sigma_z^{A_I} I^{A_O} + I^{A_I} \sigma_z^{A_O} + \sigma_z^{A_I} \sigma_z^{A_O}\right)\,,\\
M^{A_I A_O}_{1|0}&=&\frac{1}{4}\left( I^{\otimes 2} - \sigma_z^{A_I} I^{A_O} - I^{A_I} \sigma_z^{A_O} + \sigma_z^{A_I} \sigma_z^{A_O}\right)\,,\\
M^{A_I A_O}_{0|1}&=&\frac{1}{4}\left( I^{\otimes 2} + \sigma_z^{A_I} I^{A_O} - I^{A_I} \sigma_z^{A_O} - \sigma_z^{A_I} \sigma_z^{A_O}\right)\,,\\
M^{A_I A_O}_{1|1}&=&\frac{1}{4}\left( I^{\otimes 2} - \sigma_z^{A_I} I^{A_O} + I^{A_I} \sigma_z^{A_O} - \sigma_z^{A_I} \sigma_z^{A_O}\right)\,,
\end{eqnarray}
\begin{eqnarray}
M^{B_I B_O}_{0|0}&=&\frac{1}{4}\left( I^{\otimes 2} + \frac{1}{\sqrt{2}} M^{B_I B_O}_{+++} + I^{B_I} \sigma_x^{B_O}\right)\,,\\
M^{B_I B_O}_{1|0}&=&\frac{1}{4}\left( I^{\otimes 2} - \frac{1}{\sqrt{2}} M^{B_I B_O}_{-+-} + I^{B_I} \sigma_x^{B_O}\right)\,,\\
M^{B_I B_O}_{0|1}&=&\frac{1}{4}\left( I^{\otimes 2}  + \frac{1}{\sqrt{2}} M^{B_I B_O}_{+--} -I^{B_I} \sigma_x^{B_O}\right)\,,\\
M^{B_I B_O}_{1|1}&=&\frac{1}{4}\left( I^{\otimes 2} - \frac{1}{\sqrt{2}} M^{B_I B_O}_{+++} + I^{B_I} \sigma_x^{B_O}\right)\,,
\end{eqnarray}
where we have defined $M^{B_I B_O}_{ \pm \pm \pm} = \sigma_z^{B_I} I^{B_O} \pm\sigma_x^{B_I} I^{B_O}\pm \sigma_z^{B_I} \sigma_x^{B_O} \pm \sigma_x^{B_I} \sigma_x^{B_O}$.

\section{Conclusions}
To summarize, we provided a general scheme with the aim to achieve an advantage in work extraction games in the presence of indefinite causal order. For a system of four qubits, we showed that the probability that the circle qubits have a zero final energy in a single realization is equal to the success probability of the ``guess your neighbor's input''  game.
Anyway, for our scheme, the average work extracted can not be larger than the one extracted for definite causal order if the qubits are not interacting.
This suggests that, in this case, non-causal process matrices do not give a violation of the second law of thermodynamics corresponding to a definite causal order, which in general bounds the amount of extractable work.
However, it is possible to add an interaction between the qubits to get a larger average work with respect to any causal process matrix. Anyway, we note that if interactions are allowed, for a definite causal order in principle Bob (or Alice) can realize a global operation to always lower to zero the energy of the circle qubits.
Concerning the final state obtained, the scheme leads to the formation of correlations between the two couples of the two qubits in the presence of indefinite causal order.
In conclusion, we hope that our results can open a new avenue in applying indefinite causal order structures and can  inspire further investigations and applications. In particular, further investigations are needed to understand how the thermodynamic time arrow given by the second law can be affected by the causal structure.

\subsection*{Acknowledgements}
The author acknowledges financial support from the project BIRD 2021 "Correlations, dynamics and topology in long-range quantum systems" of the Department of Physics and Astronomy, University of Padova.

\appendix

\section{Upper bound}\label{appendix}

To prove the bound of Eq.~\eqref{eq bb}, we note that by using Eq.~\eqref{cond_prob} we get
\begin{eqnarray}
\nonumber p_{succ}-p_2 &=& \frac{1}{4}\text{Tr}\bigg\{\bigg(\left(M^{A_I A_O}_{0|0}-M^{A_I A_O}_{1|1}\right) \left(M^{B_I B_O}_{0|0}-M^{B_I B_O}_{1|1}\right)\\
 && - \left(M^{A_I A_O}_{0|1}-M^{A_I A_O}_{1|0}\right) \left(M^{B_I B_O}_{0|1}-M^{B_I B_O}_{1|0}\right) \bigg)W\bigg\}\,,
\end{eqnarray}
where the tensor products are implicit, from which
\begin{eqnarray}
\nonumber p_{succ}-p_2 &=& \frac{1}{4}\text{Tr}\bigg\{\bigg(\left(M^{A_I A_O}_{0}-M^{A_I A_O}_{1}\right) \left(M^{B_I B_O}_{0|1}-M^{B_I B_O}_{1|0}\right)\\
 && + \left(M^{A_I A_O}_{0|1}-M^{A_I A_O}_{1|0}\right) \left(M^{B_I B_O}_{0}-M^{B_I B_O}_{1}\right) \bigg)W\bigg\}\,,
\end{eqnarray}
where $M^{A_I A_O}_{x}=\sum_a M^{A_I A_O}_{a|x}$ and $M^{B_I B_O}_{y}=\sum_b M^{B_I B_O}_{b|y}$, which are such that $\Tr{M^{A_I A_O}_{x} M^{B_I B_O}_{y} W}=1$ for any $x$ and $y$ (see, e.g., Ref.~\cite{Oreshkov12}).
In general, the process matrix $W$ can be written as
\begin{equation}
W = \Delta(W) + \sum_{\alpha,i,j} c_{\alpha i j} \sigma^{A_I}_\alpha\sigma^{A_O}_i\sigma^{B_I}_j I^{B_O} + c'_{i \alpha j} \sigma^{A_I}_i I^{A_O}\sigma^{B_I}_\alpha\sigma^{B_O}_j\,,
\end{equation}
where $\Delta(W)= {_{A_O B_O}W}$ and $c_{\alpha i j}$ and $c'_{i \alpha j}$ are real parameters. We have that $\ParTr{A_O}{O^{A_I} I^{A_O}M^{A_I A_O}_{x}}=O^{A_I}$ for any $O^{A_I}$, and a similar equation for $M^{B_I B_O}_{y}$, so that
\begin{equation}\label{eq diff}
 p_{succ}-p_2 = \frac{1}{4}\sum_{\alpha,i,j} c_{\alpha i j} m_{\alpha i j} + c'_{i \alpha j} m'_{i \alpha j}\,,
\end{equation}
where we have defined
\begin{eqnarray}
\nonumber m_{\alpha i j } &=& \Tr{\left(M^{A_I A_O}_{0}-M^{A_I A_O}_{1}\right) \sigma^{A_I}_\alpha\sigma^{A_O}_i}\\
&& \times \Tr{ \left(M^{B_I B_O}_{0|1}-M^{B_I B_O}_{1|0}\right) \sigma^{B_I}_j I^{B_O}} \,,\\
\nonumber m'_{i \alpha j} &=& \Tr{\left(M^{A_I A_O}_{0|1}-M^{A_I A_O}_{1|0}\right) \sigma^{A_I}_i I^{A_O}}\\
&& \times \Tr{\left(M^{B_I B_O}_{0}-M^{B_I B_O}_{1}\right)\sigma^{B_I}_\alpha\sigma^{B_O}_j} \,.
\end{eqnarray}
We note that the operator $M^{A_I A_O}_{a|x}$ can be expressed as
\begin{equation}
 M^{A_I A_O}_{a|x} = \frac{q_{a|x}}{2} \bigg( I^{\otimes 2} + \vec{r}_{a|x}\cdot \vec{\sigma}^{A_I} I^{A_O} +  I^{A_I} \vec{\sigma}^{A_O}\cdot \vec{s}_{a|x} + \sum_{i,j} t^{a|x}_{ij} \sigma_i^{A_I} \sigma_j^{A_O}\bigg)\,,
\end{equation}
where $0\leq q_{a|x} \leq 1$, $\sum_a q_{a|x}=1$ and $\sum_a q_{a|x} \vec{r}_{a|x}=0$ since $\ParTr{A_O}{M^{A_I A_O}_{x}}=I^{A_I}$. Thus, we get
\begin{eqnarray}
\sum_{\alpha,i}\abs{\Tr{\left(M^{A_I A_O}_{0}-M^{A_I A_O}_{1}\right) \sigma^{A_I}_\alpha\sigma^{A_O}_i}}^2 &=& 4 \left(  ||\vec{s}_{0}- \vec{s}_1||^2 +  \sum_{i,j}\abs{t^0_{ij}-t^1_{ij}}^2\right)\\
 &\leq& 16\,,
\end{eqnarray}
where $\vec{s}_x = \sum_a q_{a|x} \vec{s}_{a|x}$ and $t^x_{ij} = \sum_a q_{a|x}t^{a|x}_{ij}$, and
\begin{eqnarray}
\sum_{i}\abs{\Tr{\left(M^{A_I A_O}_{0|1}-M^{A_I A_O}_{1|0}\right) \sigma^{A_I}_i I^{A_O}}}^2 &=& 4 ||q_{0|1} \vec{r}_{0|1}-q_{1|0} \vec{r}_{1|0}||^2 \\
 &\leq & 4\,.
\end{eqnarray}
Thus, we define the vectors $\vec{m}$ and $\vec{m}'$ with components $m_{\alpha i j}$ and $m'_{i \alpha j}$, respectively, and we get $||\vec{m}|| \leq 8$ and $||\vec{m}'|| \leq 8$.
From Eq.~\eqref{eq diff}, we get
\begin{equation}
|p_{succ}-p_2| = \frac{1}{4} \left|\vec{c}\cdot \vec{m} + \vec{c}'\cdot\vec{m}' \right|\leq \frac{1}{4}\left( |\vec{c}\cdot \vec{m}| + |\vec{c}'\cdot\vec{m}'| \right)\,,
\end{equation}
where we have defined the vectors $\vec{c}$ and $\vec{c}'$ with components $c_{\alpha i j}$ and $c'_{i \alpha j}$, respectively.
To find an upper bound of $|\vec{c}\cdot \vec{m}|+|\vec{c}'\cdot\vec{m}'|$, we consider the case where $\vec{c}$ has only one non-zero component, which is $c/4$, and $\vec{c}'$ has only one non-zero component, which is $c'/4$.
Since $W\geq 0$, there are two cases: or $|c|\leq |\cos\theta|$ and $|c'|\leq |\sin\theta|$ for a certain $\theta$, e.g., for process matrices as
\begin{equation}\label{eq. W1}
W=\frac{1}{4}\left(I^{\otimes 4} + c\sigma_z^{A_I}\sigma_z^{A_O}\sigma_z^{B_I}I^{B_O}+c'\sigma_z^{A_I}I^{A_O}\sigma_x^{B_I}\sigma_x^{B_O}\right)\,,
\end{equation}
or $|c|+|c'|\leq 1$, e.g., for process matrices as
\begin{equation}\label{eq. W2}
W=\frac{1}{4}\left(I^{\otimes 4} + c\sigma_z^{A_I}\sigma_z^{A_O}\sigma_z^{B_I}I^{B_O}+c'\sigma_z^{A_I}I^{A_O}\sigma_z^{B_I}\sigma_z^{B_O}\right)\,.
\end{equation}
It is easy to see that in both cases, $|\vec{c}\cdot \vec{m}|+|\vec{c}'\cdot\vec{m}'|\leq 2$, from which $|p_{succ}-p_2|\leq 1/2$. 


\end{document}